\documentclass[preprint,epsfig,axodraw,nofootinbib]{revtex4-1}
\usepackage{amssymb}
\usepackage{amsmath}
\usepackage{graphicx}
\usepackage{fancyhdr}
\usepackage{epsfig}
\usepackage{color}
\usepackage[dvipsnames,usenames]{xcolor}
\usepackage{hyperref}
\usepackage{soul}
\usepackage{ulem}

\hypersetup{
    unicode=false,          
    pdftoolbar=true,        
    pdfmenubar=true,        
    pdffitwindow=false,     
    pdfauthor={William},     
    colorlinks=true,       
    linkcolor=blue,          
    citecolor=red,        
    urlcolor=blue
}
\def\br{\begin{eqnarray}}
\def\er{\end{eqnarray}}
\def\be{\begin{equation}}
\def\ee{\end{equation}}

\def\to{\rightarrow}

\def\({\left(}
\def\){\right)}

\def\lesssim{\mathrel{\hbox{\rlap{\hbox{\lower4pt\hbox{$\sim$}}}\hbox{$<$}}}}
\def\gtrsim{\mathrel{\hbox{\rlap{\hbox{\lower4pt\hbox{$\sim$}}}\hbox{$>$}}}}

\begin{document}


\title{Fractionary Charged Particles Confronting Lepton Flavor Violation and the Muon's Anomalous Magnetic Moment
}

\author{Elmer Ramirez Barreto}
\email{elmer.ramirez@upch.pe}
\affiliation{ Departamento de Ciencias Exactas, Facultad de Ciencias y Filosofia, Universidad Peruana Cayetano Heredia, Av. Honorio Delgado 430, Lima 31, Peru.}

\author{Alex G. Dias}
\email{alex.dias@ufabc.edu.br}
\affiliation{Centro de Ci\^encias Naturais e Humanas, Universidade Federal do ABC,\\
09210-580, Santo Andr\'e-SP, Brasil}

\date{\today}

\begin{abstract}

In light of the recent result published by the Fermilab Muon $(g-2)$ experiment, we investigate a simple model that includes particles of
fractional electric charges: a colour-singlet fermion and a scalar with charges $2/3e$ and $1/3e$, respectively. The impact of these particles
on the anomalous muon's magnetic moment is examined, particularly the restrictions on their Yukawa couplings with the light leptons. Given that
lepton flavor violation processes impose stringent constraints on certain scenarios beyond the Standard Model, we asses the one-loop contribution
of the new particles to $(g-2)$ in order to identify regions in the parameter space consistent with the Fermilab results and compatible with 
the current and projected limits on the branching ratio $Br(\mu \rightarrow e \gamma)$. Taking into account the current lower bound for the
masses of fractionary charged particles, which is around 634 GeV, we show that the mass of the scalar particle with fractional charge must 
exceed 1 TeV. In particular, 
we present some estimatives for double production of the colour-singlet fermion at the 14 TeV LHC. Finally, we also study the validity of our model in 
light of the QCD lattice results on the muon  $(g-2)$.

\end{abstract}

\maketitle

\section{Introduction}

Recently,  the measurement of the anomalous muon's magnetic moment, $a_{\mu}= \frac{1}{2}(g- 2)_{\mu}$,  
have been updated  by the Fermilab Muon g-2 
collaboration. The new measured value, $a_{\mu}^{FNAL} = (116 592 059 \pm 22) \times 10^{-11}$ \cite{Muong-2:2023cdq} 
is in agreement with the previous results from Fermilab Muon g-2 
collaboration $a_{\mu}^{FNAL} = (116 592 040  \pm 54) \times 10^{-11}$ \cite{Abi:2021gix} and 
the Brookhaven E821 muon g-2 experiment,  $a_{\mu}^{E821} = (116592089 \pm 63)\times 10^{-11}$ 
\cite{Bennett:2004pv, Bennett:2006fi}. 

With the combination of these results, there is now  a $5.1\sigma$ deviation from the Standard Model (SM) prediction, 
$a_{\mu}^{SM} =(116591810 \pm 43)\times 10^{-11}$ \cite{Aoyama:2020ynm,Aoyama:2012wk,Aoyama:2019ryr,Czarnecki:2002nt,Gnendiger:2013pva,Davier:2017zfy,Keshavarzi:2018mgv,Colangelo:2018mtw,Hoferichter:2019mqg,Davier:2019can,Keshavarzi:2019abf,Kurz:2014wya,Melnikov:2003xd,Masjuan:2017tvw,Colangelo:2017fiz,Hoferichter:2018kwz,Gerardin:2019vio,Bijnens:2019ghy,Colangelo:2019uex,Blum:2019ugy,Colangelo:2014qya}, 
given by the discrepancy $\Delta a_{\mu}=a_{\mu}^{EXP}-a_{\mu}^{SM}=(250 \pm 48)\times 10^{-11}$. 
This confirms the previous  $4.2\sigma$ and $3.7\sigma$ deviations for $\Delta a_{\mu}$ obtained by Fermilab and the BNL Collaboration \cite{Muong-2:2006rrc}, but with higher
statistics. If this discrepancy is not due to unknown theoretical and experimental uncertainties, it raises the possibility that there is a new physics, at an energy scale not so far above the electroweak scale, manifesting through new 
particles interacting with the leptons. 

On the other hand, any assumption of the existence of new particles interacting with leptons must consider the stringent constraints imposed by lepton flavour violation (LFV) processes. A cruxial constraint arise from the absence of observed muon decay into electron and a photon, whose branching ratio is strongly limited for the case of the anti-muon  according to $Br(\mu^+ \rightarrow e^+\, \gamma) < 4.2 \times 10^{-13}$ \cite{ParticleDataGroup:2020ssz},  with a projected limit expected to reach $6 \times 10^{-14}$~\cite{Baldini:2018nnn}. Limits on LFV like this play an important role in constraining the masses and couplings of these new particles, thereby pointing on whether any potential new physics may appear just above the electroweak scale. Examples of other hypothetical LFV processes involving charged leptons with restrictive upper bounds on the branching ratios are the three body decays, with $Br(\mu^+ \rightarrow e^+\,\gamma\,\gamma) < 7.2 \times 10^{-11}$ \cite{Bolton:1988af}, $Br(\mu^+ \rightarrow 
e^+\, e^+\, e^-) < 1.0 \times 10^{-12}$ \cite{SINDRUM:1987nra} (with planed future sensitivity of $2.0 \times 10^{-15}$ \cite{Mu3e:2020gyw}) and the muon-electron conversion, with $Br(\mu^-\, {\cal N} \rightarrow e^-\, {\cal N}) $ varing from $7.0 \times 10^{-13}$ to $7.0 \times 10^{-11}$, depending on the nucleus ${\cal N}$ \cite{Badertscher:1980bt,SINDRUMII:1993gxf,SINDRUMII:2006dvw,SINDRUMII:1996fti} (with planed future sensitivities to reach $\sim 10^{-17}$ \cite{Mu2e:2014fns,COMET:2018auw}). 

Different models have been developed to explain the anomalous magnetic moment of the muon, i. e. to eliminate the large discrepancy between
the measured anomalous magnetic moment of the muon and its predicted value, keeping at the same time consistency  with the
current constraints imposed by charged LFV processes and other observables  \cite{Lindner:2016bgg,Aoyama:2020ynm,Keshavarzi:2021eqa}.
Examples are the supersymmetric models \cite{Abdughani:2019wai,Baer:2021aax,Cao:2019evo,Endo:2021zal,Baum:2021qzx,Ahmed:2021htr,
VanBeekveld:2021tgn,Abdughani:2021pdc,Han:2021ify,Gu:2021mjd,Cao:2021tuh,Altmannshofer:2021hfu,Li:2021pnt,Zhao:2022pnv,Jia:2023xpx}, 
extra dimensions-grand unification frameworks \cite{Anchordoqui:2021llp,Wang:2021bcx,Aboubrahim:2021phn,Megias:2017dzd,Brune:2022gnu,Li:2023tlk},  
models with an extended gauge group or extended scalar sector \cite{Athron:2021iuf, Li:2021lnz, Alvarado:2021nxy, Yang:2021duj, Ma:2021fre,
Dinh:2020pqn, Hue:2021xap,Cen:2021ryk, Dutta:2021afo, Chun:2021rtk, Han:2021gfu, Das:2021zea, Arcadi:2021yyr, Jueid:2021avn, Hernandez:2021iss,
Zhou:2021vnf, Iguro:2023jkf,Jia:2021mwk,Hooper:2023xnx,Ghosh:2023dgk}, and within the Standard 
Model gauge group  the inclusion of  vector-like leptons, leptoquarks and extra scalars  \cite{Poh:2017tfo, deJesus:2020upp, Dermisek:2021ajd,
Bai:2021bau, Chakrabarty:2020jro, Keung:2021rps, Nomura:2021oeu, Zhang:2021dgl, FileviezPerez:2021lkq,De:2021crr,Crivellin:2018qmi,
Athron:2022qpo,deJesus:2023som}. 

In this work, considering the SM gauge group, we investigate to what extent specific leptons and scalars with fractional charges can account for the anomalous magnetic moment of the muon, while satisfying the current limit on the branching ratio ${Br}(\mu^+ \rightarrow e^+ \,\gamma)$. For the model we consider, this is actually the most restrictive branching ratio for LFV processes. The muon three body decays turns out to be more suppressed and, like the muon-electron conversions, have upper bounds above that of the ${Br}(\mu^+ \rightarrow e^+ \,\gamma)$. The possibility of new color singlets fermions, which we will also denote by  leptons for short, and scalars with fractional charges has already been considered in the literature, including the Standard Model framework \cite{Shrock:2008sb, Langacker:2011db}, 3-3-1 models \cite{RamirezBarreto:2019bpx, Hue:2015mna} in models of grand unification \cite{Li:1981un} and also in models colorless bound systems involving new  quarks \cite{Khlopov:1981wm}. Particles 
with 
non-conventional electric charge could also originate from a mechanism like the one proposed in~\cite{Holdom:1985ag}, where  a  gauge boson associated with a U$(1)^\prime$ symmetry kinetically mixes with the SM hypercharge gauge boson SM, allowing a fermion with only U$(1)^\prime$ charge to couple with the photon. Experimentally, fractionally charged particles were investigated through Drell-Yan pair production, leading to the exclusion of particles with charges $1/3e$ 
 and $2/3e$ for masses below 200 GeV and 480 GeV,  respectively~\cite{CMS:2012xi,CMS:2013czn}. Recent analysis, utilizing data from proton-proton collisions at a center-of-mass of 13 TeV, have stablished exclusion limits for mass up to $636$ GeV and charges above $1/2e$~\cite{CMS:2022mfm}.

In this context, we consider a model with renormalizable interactions of particles with fractional electric and look for regions of the parameter space, out of their masses and coupling constants, that are consistent with the current experimental results as well as the projected ones. Our results on leptons with fractional charges are complementary  the searches in high energy colliders \cite{CMS:2012xi,CMS:2013czn, Fairbairn:2006gg, Burdin:2014xma, Golling:2016thc, Lee:2018pag, Ball:2020dnx}.

It has to be pointed out that recent results coming from  high-precision QCD lattice simulations show agreement with the experimental
measurements of the muon's anomalous magnetic moment, reducing the $\Delta a_{\mu}$ deviation to $1.5\, \sigma$ 
\cite{Borsanyi:2020mff,Ce:2022kxy}. As these results could be in conflict with the $e^+ e^-$ data \cite{Colangelo:2022vok}, various theory 
and lattice groups are expected to present updated results in order to confirm it. For the purposes of our study, we will assume that the
anomaly exists with the  $5.1\,\sigma$ deviation.

In the next section we construct a simple model of leptons and scalars with fractional electric charges within the SM symmetry group. Following this, we present our results, discussions and conclusions.

\section{The simplest model of leptons with fractional charge} 

The simplest renormalizable interaction model of a lepton with fractional electric charge coupled to the SM fermion is built by introducing  a vector-like fermion 
\begin{equation}
    \mathcal{E}^{(n)}\sim\left(\mathbf{1,\,}n\right),
\end{equation}
and the scalar  
\begin{equation}
    h^{(1-n)}\sim(\mathbf{1},1-n),
\end{equation}
which are both singlets of the $\mathrm{SU}\left(2\right)_{L}$ symmetry of the SM and have electric charges $n$ and $1-n$, respectively. The numbers in parenthesis represent the field transformations under the $\mathrm{SU}\left(2\right)_{L}$ and hypercharge $\mathrm{U}\left(1\right)_{Y}$ symmetries of the SM. These  fields are supposed to couple with the SM right-handed lepton singlets $\ell_{R}\sim\left(\mathbf{1},\,-1\right)$, $\ell=e,\,\mu,\,\tau$, through the interaction term in the Lagrangian 
\begin{eqnarray}
\mathcal{L} \supset \mathcal{Y}_{\ell \mathcal{E}}\: \overline{(\ell_{  R})^c}\:\mathcal{E}^{\: n}\,h^{1-n} + {H.c.}, 
\label{lyuk}
\end{eqnarray}
with $c$ meaning charge conjugation and $\mathcal{Y}_{\ell \mathcal{E}}$ is a coupling constant, which we assume to be real. From now on we will specialize to the case $n=2/3$. 

The corresponding interaction Lagrangian in Eq. (\ref{lyuk}) is similar to the one proposed in Refs.~\cite{Zee:1980ai, Zee:1985rj}, where a single charged Higgs is included. The $\mathcal{E}$ field is called a leptonic one by the reason we assume it carries a charge of lepton number, so that the interaction Lagrangian in Eq.~(\ref{lyuk}) invariant under such symmetry. With a single $\mathcal{E}$ field it is not possible to have lepton number distinguishing the families.

\begin{figure}
	\includegraphics[width=0.75\textwidth]{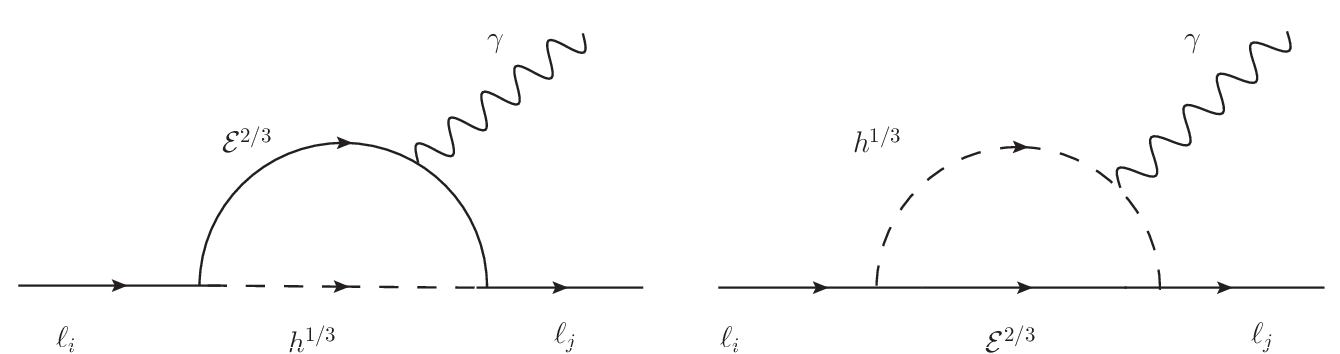}
	\caption{One-loop Feynman diagrams involving contributions from  $\mathcal{E}^{\:+ 2/3}$  and $h^{+ 1/3}$ for Br$(\mu \rightarrow e \,\, \gamma)$ and $\Delta a_{\mu}$.}	
 \label{fig1}
\end{figure}

The vector-like lepton and the charged scalar couplings with the photon field $ A_\mu$ are given by 
\begin{eqnarray}
\mathcal{L} & \supset &  - \frac{2e}{3} \overline{\mathcal{E}^{2/3}} \gamma^\mu \mathcal{E}^{2/3}A_\mu+\frac{e}{3} \left(h^{1/3\,\dagger} \partial^\mu h^{1/3} - h^{1/3} \partial^\mu h^{1/3\,\dagger} \right)A_\mu .
\label{lete}
\end{eqnarray} 
With the interactions in Eqs.~(\ref{lyuk}) and (\ref{lete}), we have new radiative corrections for the family lepton number violation muon decay, $\mu \rightarrow e \,\, \gamma$, and the anomalous magnetic moment of the muon according the Feynman diagrams in Figure \ref{fig1}. 

\subsection{Lepton flavor violation decay for the muon}
\label{mlfv}

The partial decay width of a lepton $\ell$ decaying into a lepton $\ell^\prime$ plus a photon is ~\cite{Lavoura:2003xp}  
\begin{equation}
\Gamma(\ell \to \ell^\prime\,\, \gamma) = \frac{\left( m_\ell^2 - m_{\ell^\prime}^2 \right)^3
\left( \left| \sigma_{L} \right|^2 + \left| \sigma_{R}\right|^2 \right)}
{16 \pi m_1^3}\, ,
\end{equation}
where $m_{\ell}(m_{\ell^\prime})$ is the lepton $\ell(\ell^\prime)$ mass; with $\sigma_L$ and $\sigma_R$ the form factors defined in the Appendix~\ref{app:ffac}. Thus, the ratio for the LFV process $\mu \rightarrow e \, \gamma$ is taken as 
\begin{equation}
 Br(\mu \rightarrow e \,\, \gamma)=\frac{\Gamma(\mu \rightarrow e \,\, \gamma)}{\Gamma(\mu \rightarrow e \,\, \bar{\nu_e} \,\, \nu_\mu)}.
\end{equation}
In this expression, the total width is assumed that the total width as the Standard Model one for the muon decay into an electron plus an electron anti-neutrino and a muon neutrino at the leading order, i. e. $\Gamma(\mu \rightarrow e \,\, \bar{\nu_e} \,\, \nu_\mu)=G_F^2\,m_\mu^5/(192\,\pi^3)$, with $G_F$ the Fermi coupling constant, which is much larger than $\Gamma(\mu \rightarrow e \,\, \gamma)$.

 Thus, we obtain for the branching ratio, the expression: 
 \begin{equation}
 Br(\ell_{1} \rightarrow \ell_2 \,\, \gamma)=\frac{3\,(4\,\pi)^3 \alpha_{em}\,(m_{1}^2-m_{2}^2)^3}{4\,G_{F}^2\,m_{1}^8}\,\left( \left| \sigma_{L} \right|^2 + \left| \sigma_{R} \right|^2 \right)Br(\ell_{1} \rightarrow \ell_2 \,\, \bar{\nu_2} \,\, \nu_1),
 \label{brgeral}
\end{equation}
where $\alpha_{em}$ is the electromagnetic fine-structure constant and from the experimental side, we know that $Br(\mu \rightarrow e\,\, \bar{\nu_e}\,\, \nu_\mu)= 100\,\%$, $Br(\tau \rightarrow e \,\,\bar{\nu_e} \,\,\nu_\tau)= 17,82\,\%$ and $Br(\tau \rightarrow \mu\,\, \bar{\nu_\mu} \,\,\nu_\tau)= 17,39\,\%$ \cite{ParticleDataGroup:2020ssz}.

\subsection{Muon's anomalous magnetic moment}

Loop diagrams with the particles $\mathcal{E}$ and $h$, as show in Figure \ref{fig1}, generate additional corrections to the muon anomalous magnetic moment $a_{\mu}= (g - 2)_{\mu} /2$. They give the following contribution to the $\Delta a_{\mu}$~\cite{Queiroz:2014zfa}  
\begin{eqnarray}
\Delta a_\mu(\mathcal{E},h)&=& \frac{Q_h}{8 \pi^2} \frac{m_\mu^2}{m_h^2} \int_{0}^{1}  dx \: \frac{\mathcal{Y}^{ \,2}_{\mu -\mathcal{E}} P(x)}{R^h(\lambda,\lambda^\prime, x)} 
+ \frac{Q_\mathcal{E}}{8 \pi^2} \frac{m_\mu^2}{m_h^2} \int_{0}^{1} dx \: \frac{\mathcal{Y}^{\,2}_{\mu -\mathcal{E}} P^{\prime}(x)}{R^\mathcal{E}(\lambda,\lambda^\prime, x)},
\label{detam}
\end{eqnarray}
with the functions of the mass ratios of $\mathcal{E}$, $h$ and the muon, i.e. $\epsilon = \frac{m_{\mathcal{E}}}{m_\mu}$, $\lambda = \frac{m_{\mu}}{m_h}$ and $\lambda^\prime = \frac{m_{\mathcal{E}}}{m_h}$, given by
\begin{eqnarray}
& & P(x) = x^3 - x^2 + \epsilon\,(x^2 - x), \\
& & P^{ \prime }(x) = x^2 - x^3 + \epsilon\,x^2, \\
& & R^{\:h}(\lambda, \lambda^{\prime},x) = \lambda^2\:x^2 + (1 - \lambda^2)\:x + \lambda^{ \prime \: 2}(1 - x), \\
& & R^{\:\mathcal{E}}(\lambda, \lambda^{\prime},x) = \lambda^2\:x^2 + (\lambda^{ \prime \: 2} - \lambda^2)\:x + (1 - x).
\end{eqnarray}
In Eq.~(\ref{detam}) $Q_{\mathcal{E}}$ and $Q_{h}$ represents the electric charge of $\mathcal{E}$ and $h$, respectively.

\section{Results}

Let us discuss the  potential phenomenological implications of our model on the observable quantities. We will consider the experimental constraints for $\Delta a_{\mu}$ from the Muon Collaboration at Fermilab, as well as the current experimental limit for the LFV branching ratio involving the muon \cite{ParticleDataGroup:2020ssz}: $Br(\mu \rightarrow e \,\, \gamma) < 4.2 \times 10^{-13}$ and the projected value \cite{Baldini:2018nnn} $Br(\mu \rightarrow e \,\, \gamma) < 6 \times 10^{-14}$. 

By utilizing the one-loop analytic expressions in Eqs. (\ref{brgeral}) and (\ref{detam}), which encompass the contributions of the new 
particles to the  muon  magnetic moment  and the branching  ratios, we explore the parameter space in order to identify points that
simultaneously satisfy  the muon $(g -2)$ anomaly and adhere to the constraints imposed by $Br(\mu \rightarrow e \,\, \gamma)$. Thus, 
the relevant input parameters in our investigation  are the masses $M_{\mathcal{E}}$, $M_{h}$ and the Yukawa couplings
$\mathcal{Y}_{\mu -\mathcal{E}}$ and $\mathcal{Y}_{e-\mathcal{E}}$. 

To facilitate our analysis, we impose limitations on the masses of the new exotic particles and Yukawa couplings based on collider searches
and the electron $(g -2)$. In fact, for the electron there are conflicting values for its discrepancy $\Delta a_{e}$ due the inconsistency
from two different measurements of the fine-structure constant, $\alpha_{em}$, entering in the theoretical value for $a_e^{SM}$. Taking into
account the $a_e^{EXP}$ value  \cite{Hanneke:2008tm,Hanneke:2010au}, we find in the literature:   
$\Delta a_e = a_e^{\mathrm{exp}}-a_e^{\mathrm{SM}}=(-8.7\pm 3.6)\times 10^{-13}$, with $\alpha_{em}$ determined from
$^{133}$Cs~\cite{Parker:2018vye}; and $\Delta a_e = a_e^{\mathrm{exp}}-a_e^{\mathrm{SM}}=(4.8\pm 3.0)\times 10^{-13}$, with 
$\alpha_{em}$ obtained using $^{87}$Rb~\cite{Morel:2020dww}. Given that new measurements of $\alpha_{em}$ are needed to settle this
inconsistency, we consider a set of parameters yielding $\Delta a_{e}(\mathcal{E},h)<10^{-13}$. This is done constraining the 
values for $\mathcal{Y}_{e -\mathcal{E}}$ appropriately. We fix initially the mass of the exotic lepton $\mathcal{E}^{\:+2/3}$ to be
650 GeV, in accordance with the experimental limits from the LHC \cite{CMS:2022mfm}. Finally, our choice of 
the $\mathcal{Y}_{l -\mathcal{E}}$ couplings also takes into account the values obtained for $m_{h^{1/3}}$  within the energy range of
the LHC, while avoiding a excessive fine-tuning.

\begin{figure}[h]
	\includegraphics[width=0.48\textwidth]{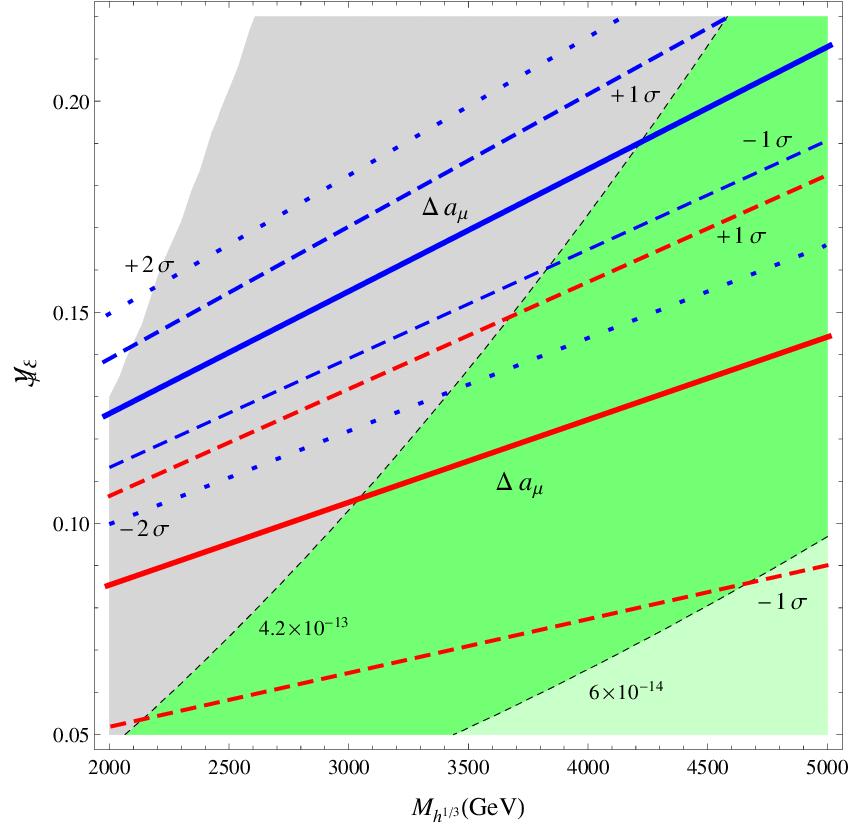}
	\includegraphics[width=0.48\textwidth]{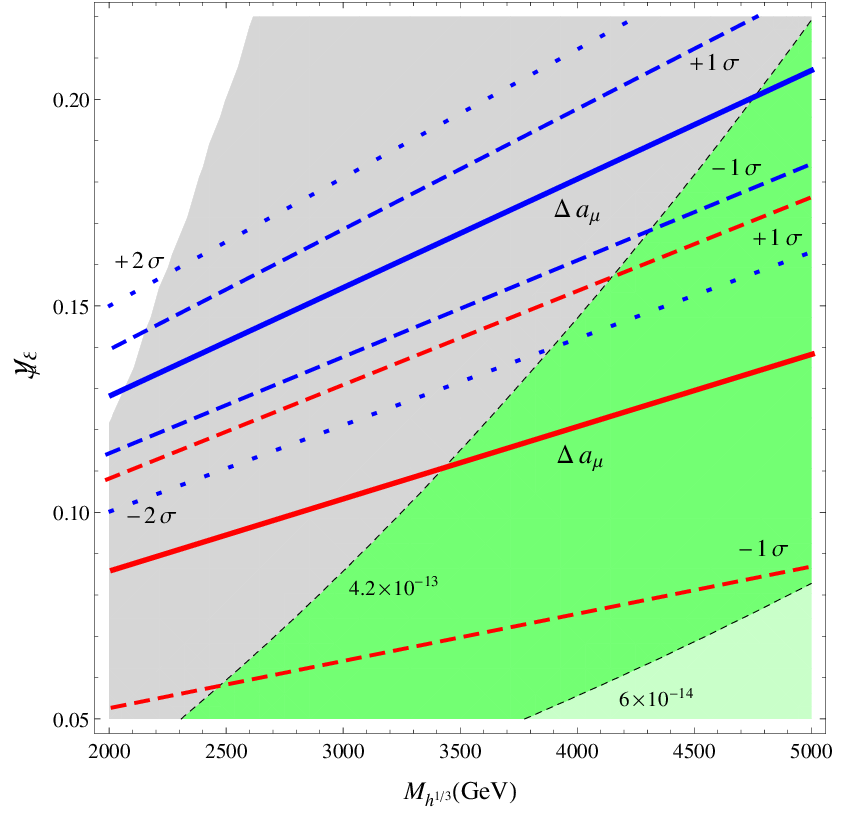}
	\caption{Contour lines representing the current and projected limit for $Br(\mu \rightarrow e \,\, \gamma)$ for $M_{\mathcal{E}}=650$ GeV (left)
	 and $M_{\mathcal{E}}=800$ GeV (right) in the 	($M_{h^{1/3}} ,\,\mathcal{Y}_{\mu -\mathcal{E}}$) plane, with $\mathcal{Y}_{e -\mathcal{E}}=10^{-6}$.
	 The blue solid line represents points in the parameter space that  match the anomalous $\Delta a _{\mu}$, while the dashed lines define 
	 experimental $1\sigma$ and $2\sigma$ range in the same space. The gray zone contains values that do not satisfy the current limit for $Br(\mu \rightarrow e \,\, \gamma)$. 
	The red solid line represents points in the parameter space that  match $\Delta a _{\mu}$ involving the lattice QCD contributions \cite{Borsanyi:2020mff,Ce:2022kxy}, with the dashed line indicating 
	 experimental $1\sigma$ range.}
	\label{fig2}       
\end{figure}

We now show in Figure \ref{fig2}, our numerical results  fixing $\mathcal{Y}_{e -\mathcal{E}}=10^{-6}$. In the left panel, for $M_{\mathcal{E}}= 650$ GeV, the green and soft green regions contains points ($M_{h}$, $\mathcal{Y}_{\mu -\mathcal{E}}$) compatible with the current and projected bounds for
$Br(\mu \rightarrow e \,\, \gamma)$, while the gray zone represent an exclusion zone for this observable. 
In addition, in the same plot, we show values for $M_{h}$ and $\mathcal{Y}_{\mu -\mathcal{E}}$ that matches the
$\Delta a_{\mu}$ data, represented by the blue line  and their respective $1 \sigma$  and $2 \sigma$ ranges. Thus, for $m_h  \geq 4.2$ TeV, and with $\mathcal{Y}_{\mu -\mathcal{E}} \geq 0.19$ the contributions of the new particles explains $\Delta a_{\mu}$ and respect the limits for $Br(\mu \rightarrow e \,\, \gamma)$. If we take $M_{\mathcal{E}}= 800$ GeV (right  panel), we have $m_h \geq 4.7$ TeV, and with $\mathcal{Y}_{\mu -\mathcal{E}} \geq 0.20$
in order to verify both observables. On the other hand, if we consider the region defined by the - 2$\sigma$ deviation, 
we identified $m_h \geq 3.2$ TeV and $\mathcal{Y}_{\mu -\mathcal{E}} \geq 0.12$ for $M_{\mathcal{E}}= 650$ GeV,  and 
$m_h \geq 3.6$ TeV and $\mathcal{Y}_{\mu -\mathcal{E}} \geq 0.12$ for $M_{\mathcal{E}}= 800$ GeV.

\begin{figure}[h]
	\includegraphics[width=0.48\textwidth]{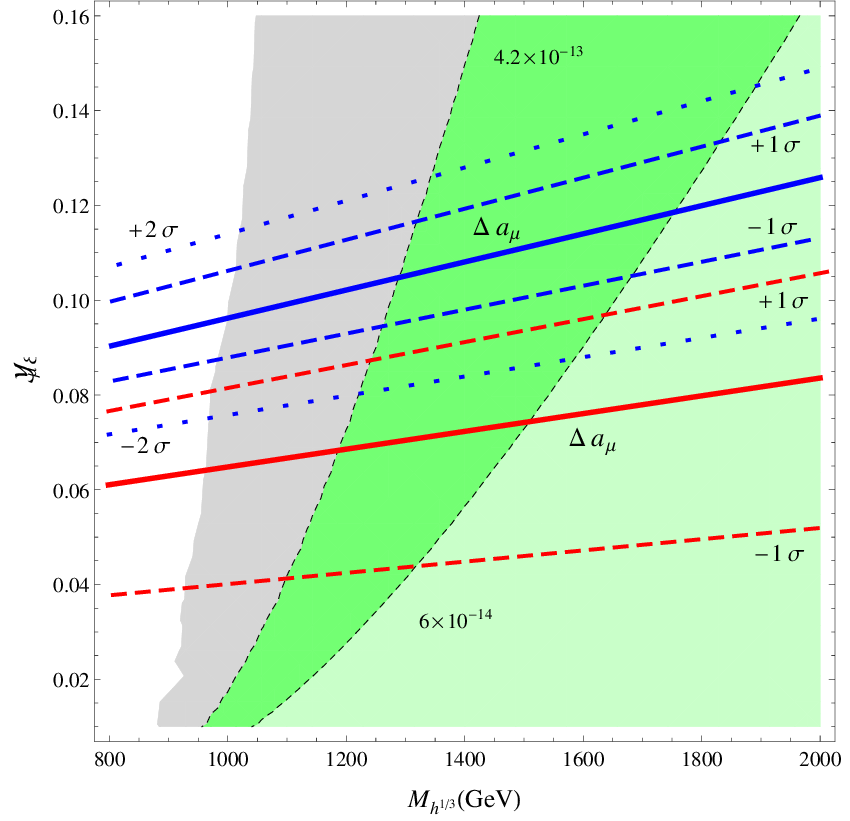}
	\includegraphics[width=0.48\textwidth]{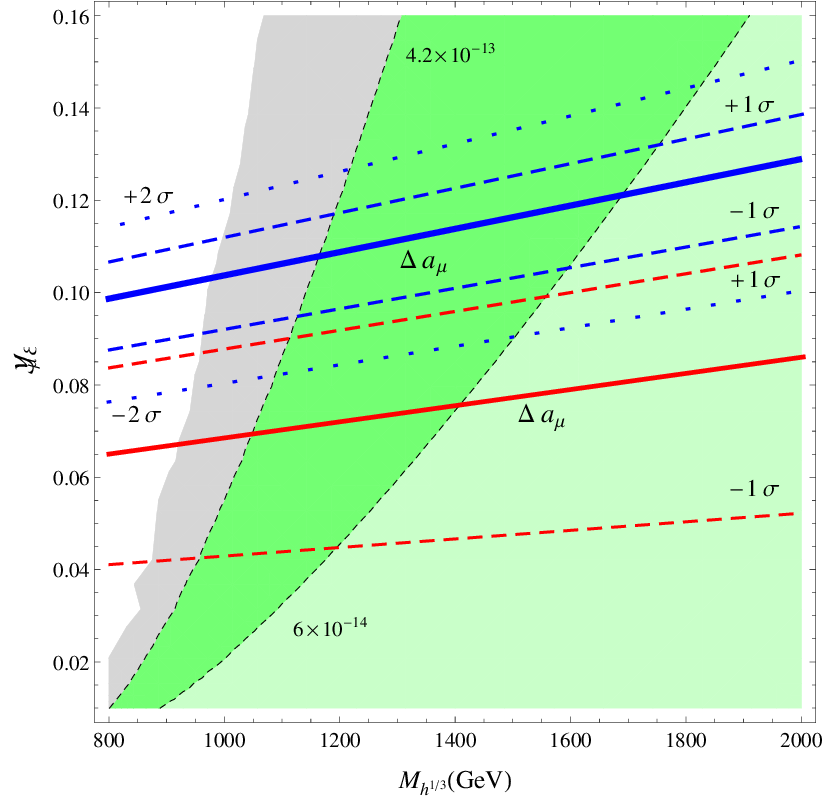}
	\caption{Contour lines representing the current and projected limit for $Br(\mu \rightarrow e \,\, \gamma)$ for $M_{\mathcal{E}}=650$ GeV (left)
	 and $M_{\mathcal{E}}=800$ GeV (right) in the 	($M_{h^{1/3}} ,\,\mathcal{Y}_{\mu -\mathcal{E}}$) plane, with $\mathcal{Y}_{e -\mathcal{E}}=10^{-7}$.
	 The blue solid line represents points in the parameter space that  match the anomalous $\Delta a _{\mu}$, while the dashed lines define 
	 experimental $1\sigma$ and $2\sigma$ range in the same space. The gray zone contains values that do not satisfy the current limit for Br. 
	The red solid line represents points in the parameter space that  match $\Delta a _{\mu}$ involving the lattice QCD contributions with the dashed line indicating 
	 experimental $1\sigma$ range.}
	\label{fig3}       
\end{figure}

In Figure \ref{fig3}, we present a new scenario with $\mathcal{Y}_{e -\mathcal{E}}=10^{-7}$. The left panel, again for $M_{\mathcal{E}}= 650$ GeV,
shows that  for  $m_h  \geq 1.25$ TeV, and with $\mathcal{Y}_{\mu -\mathcal{E}} \geq 0.10$
it is posible to explain $\Delta a_{\mu}$ and respect the $Br(\mu \rightarrow e \,\, \gamma)$ limits. In addition, for $M_{\mathcal{E}}= 800$ GeV (right panel),
the limits are increased, so $m_h  \geq 1.1$ TeV and $\mathcal{Y}_{\mu -\mathcal{E}} \geq 0.11$ allowing us to explain $\Delta a_{\mu}$ and
the respective current constrain for $Br(\mu \rightarrow e \,\, \gamma)$. If we consider again the region defined by the - 2$\sigma$ deviation, we identified  $m_h \geq 1.2$ TeV and $\mathcal{Y}_{\mu -\mathcal{E}} \geq 0.07$ for $M_{\mathcal{E}}= 650$ GeV,  and $m_h \geq 1.05$ TeV and $\mathcal{Y}_{\mu -\mathcal{E}} \geq 0.075$ for $M_{\mathcal{E}}= 800$ GeV. 

\begin{figure}
	\includegraphics[width=0.7\textwidth]{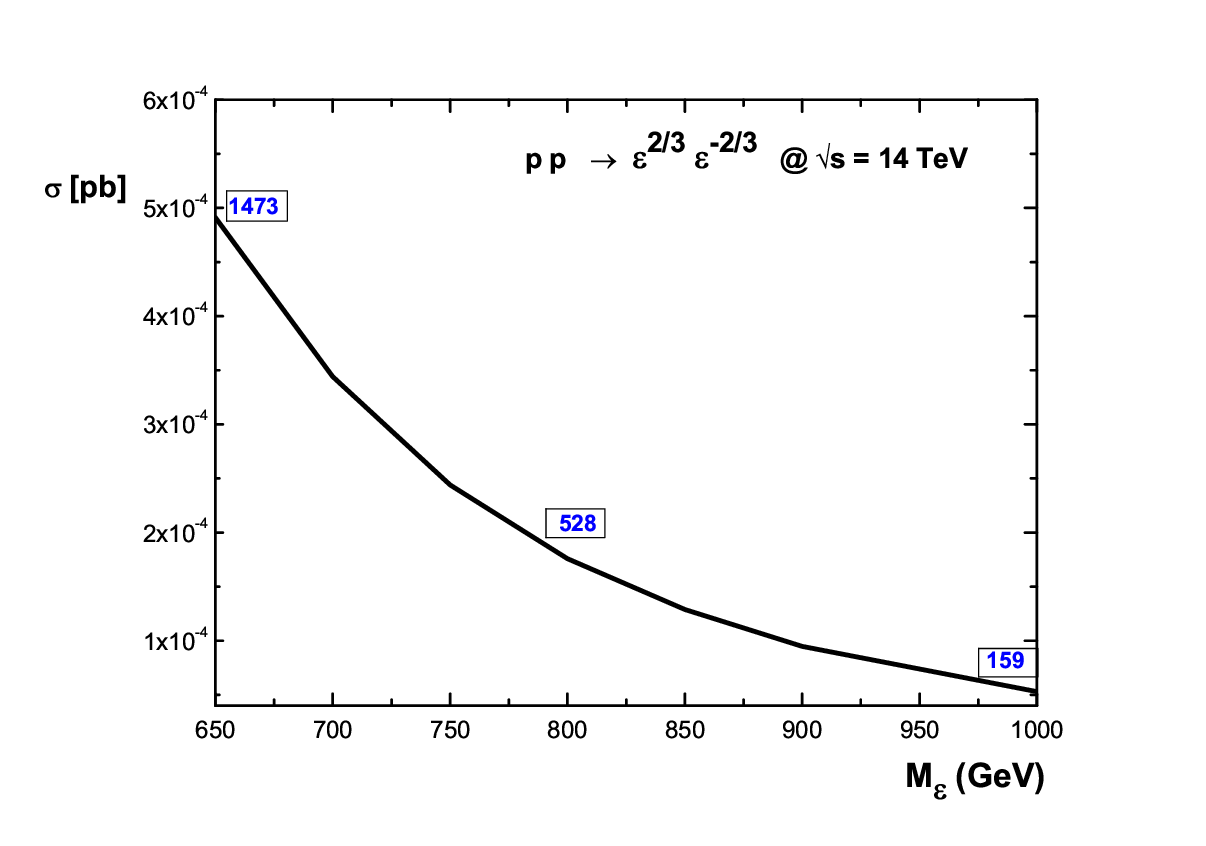}
	\caption{Cross section for the $\mathcal{E}^{\:+ 2/3}$ pair production at the 14 TeV LHC. The numbers over the curve are the expected number of events taking into account the integrated luminosity of    3 ab$^{-1}$ of integrated luminosity at the LHC run III}	
 \label{fig4}
\end{figure}

For this last scenario, the considered exotic lepton masses and the exotic scalar mass are within the range of energies reached by the LHC so 
that these particles could be produced  through the Drell-Yan processes $p p \rightarrow \mathcal{E}^{\: 2/3} \mathcal{E}^{\:- 2/3} $ and
$p p \rightarrow h^{+ 1/3} h^{- 1/3}$ at the next high-luminosity run of the LHC. For example, by simulating the $\mathcal{E}^{\: 2/3}$ pair
production  with $\sqrt{s}=14$ TeV for $M_{\mathcal{E}}= 650$ GeV using the CalcHep package \cite{Belyaev:2012qa}, we found a cross-section
$\sigma=4.9\times10^{-4}$ pb. If we take into account the projected luminosity for the LHC run III, $\mathcal{L} = 3$ab$^{-1}$, will be 
produced around 1473 events. In  Figure \ref{fig4} we show the $\mathcal{E}^{\: 2/3}$ pair production cross-section and the expected number of events for some mass benchmarks. We can expect that for masses as in the former scenario, $\sim 4$ TeV, there might be a reasonable discovery potential for these particles at the 
future higher colliders, such as High-Energy LHC.

Finally, as already mentioned in the introduction, the latest lattice results predict a larger value of muon $(g -2)$ bringing
it closer to experimental value. In this sense, our model remains consistent with such result for $\Delta a _{\mu}$, respecting the limits for $Br(\mu \rightarrow e \,\, \gamma)$. Thus, the contributions of the exotic particles in that scenario are represented by red lines in the figures 2 and 3, where for $\mathcal{Y}_{e -\mathcal{E}}=10^{-6}$ (figure 2) the lattice QCD results involves $m_h \gtrsim 3$ TeV  and $\mathcal{Y}_{\mu -\mathcal{E}} \gtrsim 0.10$ for both fixed $M_{\mathcal{E}}$. On the other hand, for $\mathcal{Y}_{e -\mathcal{E}}=10^{-7}$ (figure 3), we identify  $m_h  \gtrsim 1.5$ TeV and $\mathcal{Y}_{\mu -\mathcal{E}} \gtrsim 0.06$ compatible with the lattice results and the respective current constrain for $Br(\mu \rightarrow e \,\, \gamma)$.

\section{Conclusions and final remarks}

In this work, we have demonstrated that the inclusion of a vector-like lepton $\mathcal{E}$ and a scalar $h$ with exotic electric charges allow us predict significant contributions to $\Delta a _{\mu}$. This enable us to explain the muon $(g -2)$ anomaly while also respecting the experimental constrains on $Br(\mu \rightarrow e \,\, \gamma)$.

By considering the contributions of such exotic particles at the one-loop level for $\Delta a _{\mu}$, and taking into account the current limits on particles with fractional electric charges, we explore the parameter space defined by $m_{h^{1/3}}$ and  $\mathcal{Y}_{\mu -\mathcal{E}}$. Through our numerical analysis, we have identified the regions of mass and Yukawa couplings in the parameter space able to explain the muon $(g -2)$ anomaly while satisfying phenomenological constraints for the charged lepton flavor-violating decay $\mu \rightarrow e \,\, \gamma$. In two benchmark scenarios, with the vector-like lepton mass $m_{\mathcal{E}}$ fixed at $650$ and $800$ GeV, we have found masses around the TeV scale for the exotic Higgs and couplings that account for the muon $(g -2)$ anomaly while satisfying the branching ratio constraint. Furthermore, if we consider a conservative scenario, which includes lattice QCD results, our model remains capable of explaining both results. From the phenomenological 
standpoint, we explore the production of the exotic leptons through the Drell-Yan processes $p p \rightarrow \mathcal{E}^{\: 2/3} \mathcal{E}^{\:- 2/3} $. So, taking into account the next LHC stage at  $\sqrt{s}= 14$ TeV and with a projected integrated luminosity of $\mathcal{L} = 3$ ab$^{-1}$, will be produced around 1473 events involving the exotic lepton with mass fixed at $650$ GeV.

We would like to call attention to the fact that there is a search at LHC looking for experimental evidence of particles with exotic charges and masses above electroweak energy scale~\cite{Mitsou:2021vhf}. Therefore, further phenomenological analysis will deserve attention in future studies.

\section*{Acknowledgments}

This study was financed in part by Conselho Nacional de Desenvolvimento Científico e Tecnológico (CNPq), grant 305802/2019-4 (A.G.D).

\appendix

\section{Form factors}
\label{app:ffac}

In this appendix, we show the  general expressions for the form factors used in subsection II.  Thus, we have:

\begin{equation}
 \sigma_L=\,q_{\mathcal{E}}\,\left[ c_1\,\kappa_1+ c_2\,\kappa_2+c_3\,\kappa_3 \right] + \,q_h\,\left[ c_1\,\bar{\kappa_1}+ c_2\,\bar{\kappa_2}+c_3\,\bar{\kappa_3} \right],
\end{equation}

\begin{equation}
 \sigma_R=\,q_{\mathcal{E}}\,\left[ d_1\,\kappa_1+ d_2\,\kappa_2+d_3\,\kappa_3 \right] + \,q_h\,\left[ d_1\,\overline{\kappa_1}+ d_2\,\overline{\kappa_2}+d_3\,\overline{\kappa_3} \right]
\end{equation}


where $q_{\mathcal{E}}$ and $q_h$ are the electric charges of the exotic particles, and the $c_i$ and $d_i$ coefficients are defined in function of the
scalar and pseudoscalar Yukawa couplings:

\begin{eqnarray}
     & & c_1 = (\mathcal{Y}^{\,s\,*}_{2}+\mathcal{Y}^{\,p\,*}_{2})(\mathcal{Y}^{\,s}_{1}+\mathcal{Y}^{\,p}_{1}),\,\,\,\, c_2 = (\mathcal{Y}^{\,s\,*}_{2}-\mathcal{Y}^{\,p\,*}_{2})(\mathcal{Y}^{\,s}_{1}-\mathcal{Y}^{\,p}_{1}),\nonumber\\ 
     & & c_3 = (\mathcal{Y}^{\,s\,*}_{2}+\mathcal{Y}^{\,p\,*}_{2})(\mathcal{Y}^{\,s}_{1}-\mathcal{Y}^{\,p}_{1}), \,\,\,\, d_1 = (\mathcal{Y}^{\,s\,*}_{2}-\mathcal{Y}^{\,p\,*}_{2})(\mathcal{Y}^{\,s}_{1}-\mathcal{Y}^{\,p}_{1}),\nonumber\\ 
     & & d_2 = (\mathcal{Y}^{\,s\,*}_{2}+\mathcal{Y}^{\,p\,*}_{2})(\mathcal{Y}^{\,s}_{1}+\mathcal{Y}^{\,p}_{1}),\,\,\,\, d_3 = (\mathcal{Y}^{\,s\,*}_{2}-\mathcal{Y}^{\,p\,*}_{2})(\mathcal{Y}^{\,s}_{1}+\mathcal{Y}^{\,p}_{1}).
\end{eqnarray}

The factors $\kappa_i$ and $\bar{\kappa_i}$ can be write as:

\begin{equation}
 \kappa_{1,2}=\frac{i\,m_{1,2}}{16\,\pi^2\,m_{h}^2}\,\left[\frac{x^2-5x - 2}{12\,(x-1)^3}+\frac{x\,lnx}{2\,(x-1)^4}\right],
\end{equation}

\begin{equation}
 \kappa_{3}=\frac{i\,m_{\ell}}{16\,\pi^2\,m_{h}^2}\,\left[\frac{x - 3}{2\,(x-1)^3}+\frac{lnx}{(x-1)^3}\right],
\end{equation}

and

\begin{equation}
 \overline{\kappa_{1,2}}=\frac{i\,m_{1,2}}{16\,\pi^2\,m_{h}^2}\,\left[\frac{2\,x^2+5\,x - 1}{12\,(x-1)^3}-\frac{x^2\,lnx}{2\,(x-1)^4}\right],
\end{equation}

\begin{equation}
 \overline{\kappa_{3}}=\frac{i\,m_{\ell}}{16\,\pi^2\,m_{h}^2}\,\left[\frac{x + 1}{2\,(x-1)^2}-\frac{x\,lnx}{(x-1)^3}\right].
\end{equation}

\bibliographystyle{apsrev4-1}

\bibliography{myrefs}

\end{document}